\def\lae{\mathrel{<\kern-1.0em\lower0.9ex\hbox{$\sim$}}}
\def\gae{\mathrel{>\kern-1.0em\lower0.9ex\hbox{$\sim$}}}

\documentclass[12pt,preprint2]{aastex}
\usepackage{emulateapj5}
\usepackage{onecolfloat5}

\slugcomment{Accepted for publication in ApJ Letters}

\shorttitle{RELATIVE AGES OF M87 GLOBULAR CLUSTERS}
\shortauthors{JORD\'AN ET AL.}

\begin{document}

\twocolumn[
\title{The Relative Ages of the Globular Cluster Subpopulations in M87\altaffilmark{1}}

\author{Andr\'es Jord\'an\altaffilmark{2,3}, Patrick C\^ot\'e\altaffilmark{3},
Michael J. West\altaffilmark{4} and Ronald O. Marzke\altaffilmark{5}}

\begin{abstract} 
Relative ages for the globular cluster (GC) subpopulations in the Virgo
giant elliptical galaxy M87 (NGC~4486) have been determined from
Str\"omgren photometry obtained with WFPC2 on board \textit{HST}.  Using a
variety of population synthesis models, and assuming the GC mass at the
turnover of the luminosity function is the same for both subpopulations,
differential ages have been determined from the observed magnitudes at the
turnover of the globular cluster luminosity function and from the mean
color of each subpopulation.  We measure an age difference between the two
subpopulations of $0.2 \pm 1.5$ (systematic) $\pm 2$ (random) Gyr, in the
sense that the blue GCs are formally older. Thus, to within our
measurement errors, the two subpopulations are found to be coeval.
Combined with previous spectroscopic age determinations for M87 GCs, our
results favor a picture in which the GCs associated with this galaxy are
formed at high redshift, and within a period of a few Gyr.
\end{abstract}

]

\altaffiltext{1}{Based on observations with the NASA/ESA {\it Hubble Space Telescope}
obtained at the Space Telescope Science Institute, which is operated by the Association
of Universities for Research in Astronomy, Inc., under NASA contract NAS 5-26555}
\altaffiltext{2}{Claudio Anguita Fellow}
\altaffiltext{3}{Department of Physics and Astronomy, Rutgers University, Piscataway, NJ 08854.
andresj@physics.rutgers.edu, pcote@physics.rutgers.edu}
\altaffiltext{4}{Department of Physics \& Astronomy, University of Hawaii, Hilo, HI 96720.
west@bohr.uhh.hawaii.edu}
\altaffiltext{5}{Department of Physics \& Astronomy, San Francisco State University,
1600 Holloway Avenue, San Francisco, CA 94132. marzke@quark.sfsu.edu}

\section{Introduction}

Globular clusters (GCs) are among the oldest stellar systems in the
Universe.  Their large numbers around early type galaxies, their ease of
detection and their simplicity as single-age, mono-metallic stellar
populations, make them ideal tracers of the evolutionary history of their
host galaxies. A key result that has emerged from observations of GC
systems is that a large fraction of early-type galaxies show a bimodal
distribution in broadband color (e.g. Gebhardt \& Kissler-Patig 1999).  
As broadband colors of old stellar populations are much more sensitive to
metallicity than age, this bimodality indicates the presence of two GC
subpopulations, a metal-rich and a metal-poor, but due to the
age-metallicity degeneracy (e.g. Worthey 1994) no firm conclusion can be
drawn on the ages of the two components based on broadband colors alone.
Several models have attempted to explain the bimodality in terms of a
particular galaxy formation process: major mergers of late-type galaxies
(Ashman \& Zepf 1992), two bursts of \textit{in-situ} star formation
(Forbes, Brodie \& Grillmair 1997) and hierarchical formation (C\^ot\'e,
Marzke \& West 1998). The first two models form the metal-rich GCs out of
a second burst of star formation while the third assumes that the GC
system is assembled via dissipationless mergers.

A key observational constraint on those models is the relative ages of the
GC subpopulations. Multiple metallicity populations do not necessarily
require separate bursts of star formation but may also reflect differences
in the environment where the GC subpopulations formed. Elucidating the
formation histories of the GC subpopulations on the basis of age can help
discriminate between the proposed models.

Most previous determinations of the ages of the GC subpopulations in
undisturbed early-type galaxies seem to suggest coeval subpopulations
(Kissler-Patig et~al 1998; Cohen, Blakeslee \& Ryzhov 1998; Puzia et~al.
1999; Larsen et~al. 2002), albeit usually with large uncertainties.  For
M87, there have been conflicting claims. Cohen et~al.  (1998)  obtained
Keck spectroscopy of $150$ GCs in M87 and found no sign of a variation in
age with metallicity, excluding the possibility that one population is
half as old as the other at the $99\%$ confidence level. On the other
hand, Kundu et~al. (1999) estimate using $V$ and $I$ photometry from
\textit{HST}, that the metal-rich clusters are $3-6$ Gyr younger than
their metal-poor counterparts.

We have obtained Str\"omgren photometry of the GC system in the inner part
of M87. By combining a metallicity-sensitive color index with the turnover
$u$-band luminosity of the GC luminosity function (GCLF) of the two
subpopulations, we have measured their relative age by comparing our
observations with population synthesis models. We find the two GC
subpopulations to be coeval within our measurement errors, in agreement
with the spectroscopic estimates.

\section{Observations and Data Reduction}

\renewcommand{\thefootnote}{\alph{footnote}}
\setcounter{footnote}{0}

We used WFPC2 on board {\it HST} to obtain Str\"omgren photometry (filters
F336W, F410M, F467M and F547M) of M87 (=NGC~4486), the giant elliptical
galaxy near the dynamical center of the Virgo cluster. The exposure times
were $24\times1200$ s (F336W), $16\times1200$ s (F410M), $8\times1200$ s
(F467M) and $8\times1200$ s (F547M). The field was positioned such that
the nucleus of M87 falls in the PC.  The raw data were processed with the
standard STScI pipeline using the best available calibration.

In order to detect GCs against the background of light from M87, we
constructed detection frames by subtracting a ring medianed image (Secker
1995) using radii of 5 pixels for the PC and 4 pixels for the WF.
Detection was done with SExtractor (Bertin \& Arnouts 1996), using
Gaussian kernels as filters.  All the detections were then photometered
using DAOPHOT (Stetson 1987). We used an aperture of $2$ pixels for the WF
chips and $3$ pixels for the PC. The sky was taken from an annulus between
$r_i=5$ pixels and $r_o=8$ pixels.

The lists of detected objects were then matched with a matching radius of
$0.2''$ to produce a final catalog of GC candidates. To distinguish bona
fide GCs from foreground stars and background galaxies we used a method
similar to the one employed by Kundu et~al. (1999), which uses the flux
ratio measured in apertures of $2$ pixels and $0.5$ pixels:
$r=f_{2}/f_{0.5}$. Detections were considered to be GCs if this ratio was
in the range $1.5 < r < 12$ for the PC and $1.5 < r < 8.5$ for the WF
chips. The matched list of GC candidates contained $628$ objects.

Instrumental magnitudes were then corrected for charge transfer efficiency
following the prescriptions of Whitmore, Heyer \& Casertano (1999). This
correction was very small for F547M but could be as large as $\sim 0.3$
mag for the faintest GCs in F336W. Instrumental magnitudes were
transformed to the VEGAMAG system using zeropoints taken from the HST Data
Handbook. A correction for foreground extinction was performed using the
redenning curves of Cardelli, Clayton \& Mathis (1989), with a value of
$E(B-V)=0.023$ taken from the DIRBE maps of Schlegel, Finkbeiner \& Davis
(1998). In computing the aperture corrections for a $0.5''$ radius, we
selected $\sim 60$ isolated cluster candidates with photometric errors
less than $0.05$ mag and used DAOGROW (Stetson 1990) to compute the
aperture corrections.  A final correction of $0.1$ mag was then applied to
correct to infinite aperture (Holtzman et~al. 1995).

We performed artificial star tests to determine the completeness function
of our detections. The same objects that were used to derive the aperture
corrections were used to create a PSF to model the GCs. We used DAOPHOT to
add $50$ GCs at a time to a given chip. The artificial images thus created
were then subjected to the same reduction process as the real data. We
added a total of $20,000$ objects per chip and obtained the completeness
function by fitting a function of the form $f(m|\alpha,m_l) = 0.5[1 -
\alpha (m-m_l)/\sqrt{1+\alpha^2(m-m_l)^2}]$ (Fleming et~al. 1995).  In
order to account for a dependence of the completeness function on the
value of the galaxy halo light underlying the GCs, we divided the results
from our artificial star tests according to the background level and
measured the respective completeness functions. After some
experimentation, we settled on separate completeness functions for objects
with backgrounds above or below $1$ ADU in the unfiltered image for both
the PC and WF chips. The completeness function for the WF chips were very
similar from chip to chip so we treated them as one.

\section{Analysis}
\label{sec:analysis}

To measure the relative age of the GC subpopulations we used a technique
similar to that of Puzia et~al. (1999). The idea is to estimate the
turnover of the GCLF separately for the metal rich and the metal poor
subpopulations; if the turnover reflects the same mass scale for both
subpopulations, then we can compare the observed mean colors and turnover
magnitudes with the predictions of population synthesis models to infer
the mean metallicity and age for each subpopulation. Note that the mass at
the turnover is \textit{not} required since the models can be shifted
arbitrarily in magnitude. Provided the assumption of equal masses at the
turnover is valid, the relative age follows from the models regardless of
the specific value of the GC turnover mass.

Puzia et~al. (1999) used $V$ and $I$ photometry from \textit{HST} to
implement this technique. We exploit the enhanced fading of stellar
populations with age in the $u$ band, and a more sensitive metallicity
index, to improve the sensitivity of the technique. We used two different
indices in our analysis: $s_1\equiv m_{F336W}-m_{F547M}$ and $s_2\equiv
m_{F336W}+m_{F410M} -m_{F467M}-m_{F547M}$. The latter has greater
metallicity sensitivity, but the ratio of the sensitivity to the typical
photometric error in each index is approximately the same for both.

\begin{figure}
\plotone{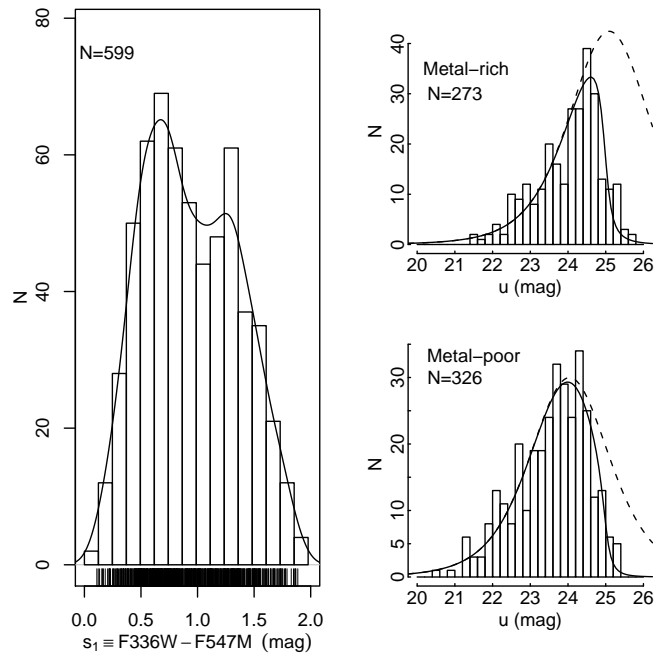}
\caption{Left: GC $s_1$ color distribution for all GCs that lie in the range $0.1<s_1<1.9$. 
The solid line
is a density estimate using a normal kernel. Right: GCLF for the metal-rich and 
metal-poor subpopulations
separated using $s_1$. The solid line is obtained by multiplying the 
best-fit $t_5$ distribution by the  completeness function, which can be compared 
directly with the observations. The dashed line is the best fit $t_5$, note the shift 
between the inferred LFs of both populations.
Note that the turnover for the metal-rich population
falls near the 50\% completeness which falls at u $\sim 25$ mag.}
\label{fig:histandlf}
\end{figure}

To reduce the background contamination we restricted the GC sample by
imposing a cut in color: we included in the analysis only those GC
candidates that lie in the range $0.1 < s_1 < 1.9$ or $0.3 < s_2 <3.0$;
this reduces the sample by $\sim 5\%$.  Due to the small size of the field
and the large number of GCs present on it, contamination in the final
sample is expected to be very small. A histogram of the $s_1$ color
distribution and a normal kernel density estimate (Silverman 1986)  is
shown in Figure~\ref{fig:histandlf}.  Bimodality is seen clearly in both
distributions and this is confirmed by a KMM test (Ashman, Bird \& Zepf
1994) which accepts the hypothesis of bimodality at $99.9\%$ confidence.

To separate the color ditribution into a metal-rich and a metal-poor
sample, and to obtain the color of the peaks of each population we used
the output of KMM, which returns a mean color for each subpopulation and a
value that divides the sample based on an \textit{a-posteriori}
likelihood. We also obtained estimates of the peaks and the dividing point
using a normal kernel density estimate which returned values in good
agreement with those from KMM. The values adopted to separate the
subpopulations are $0.97$ mag for $s_1$ and $1.5$ mag for $s_2$. The mean
colors for both subpopulations are listed under $m_{col}$ in
Table~\ref{tab:res}.

Having divided the sample into  metal-rich and  metal-poor subpopulations
we used a maximum-likelihood method 
in order to estimate the turnover of the $u$ band GCLF for 
each subpopulation.
Explicitly, if we denote the parametric representation of the intrinsic LF by 
$\Phi(u|\Theta)$, where 
$\Theta$ is the set of parameters that determine the LF, and the completeness function by 
$f(u|b)$, where $b$ is the background level,  we maximize the likelihood
$L(\Theta) = \prod_{i=1}^{n}A_i\Phi(u_i|\Theta)f(u_i|b_i)$
where $A_i$ is a normalization factor.
For $\Phi(u|\Theta)$, we used a Student $t_5$ distribution,
\( \Phi(u|\mu ,\sigma_t)=(8/3 \sqrt{5}\pi\sigma_t)[1+(u-\mu)^2/5\sigma_t^2]^{-3} \), 
which
has been shown to be a better fit to the MW GCLF than a normal distribution (Secker 1992).
We let $\sigma_t$ be free, or fixed at $\sigma_t=1.1$ mag (equivalent to 
$\sigma_g=1.4$ mag for a Gaussian distribution), which is consistent with the 
value obtained for M87 by Kundu et~al. (1999), $\sigma_g = 1.39\pm 0.06$ mag.
The fit with two free parameters confirmed that the choice of $\sigma_t=1.1$  mag is supported by our
data for both subpopulations, so in subsequent analysis we used the one-parameter fit.
Uncertainties were calculated both with a bootstrap procedure and using the maximum likelihood
surface around the minimum. 
The results for the turnovers are presented in 
Table~\ref{tab:res} under $m_{TO}$ for both GC subpopulations.
Varying the turnover estimation by letting both $\mu$ and $\sigma_t$ be free
and using a Gaussian distribution ($\sigma_g=1.4$) instead of a $t_5$ for 
$\Phi(u|\Theta)$ does not alter our 
conclusions. Likewise, disregarding the background dependence and considering
a single completeness function $f(u)$ leads to similar results. 
Possible biases introduced by the difference between the intrinsic LFs and the assumed 
models will be investigated in future work (Jord\'an et~al., in preparation).

\section{Results and Discussion}
\label{sec:resdis}

\begin{figure}
\plotone{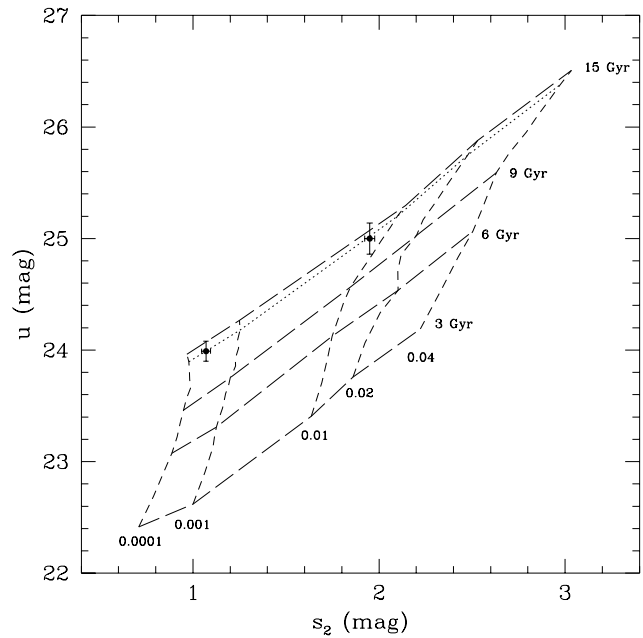}
\caption{Determination of the relative ages of the GC populations using Maraston's (1998) models
with a Gould et~al. (1997) IMF. The short dashed lines are isometallicity tracks of
$Z=0.0001,0.001,0.01,0.02,0.04$ and the long dashed lines are isochrones of $3$, $6$, $9$ and $15$ Gyr.
The points are at the estimated turnover luminosities and mean $s_2$ color for both populations; the
error bars are $1\sigma$ uncertainties.
The dotted line is the isochrone of $14$ Gyr where the blue population has been assumed to lie. The
age for the red population, and thus the relative age, is then interpolated from the grid. (Some 
isochrones have been omitted for clarity).}
\label{fig:grid}
\end{figure}

The turnover magnitudes for the metal-rich and metal-poor subpopulations
of M87 and their respective mean colors were compared to the predictions
of population synthesis models. We used the latest versions of the models
of Bruzual \& Charlot (1993), Maraston (1998)  and Worthey (1994) to
compute grids for every available metallicity. Because a particular value
for the mass at the turnover is not assumed, the \textit{absolute} ages
for the subpopulations are unknown.  Instead, the model grid was shifted
in such a way that one of the subpopulations lies on the $14$ Gyr
isochrone, which is consistent with the age of the oldest MW GCs
(VandenBerg 2000), and then the relative age was obtained by linearly
interpolating from the model grid.  Choosing another age for the oldest
isochrone does not change our results significantly. A graphical example
of the procedure can be seen in Figure~\ref{fig:grid}. For each set of
models we tried two different IMFs: a Salpeter (1955) and a Miller \&
Scalo (1979)  for the Bruzual \& Charlot (1993) and Worthey (1994) models
and a Salpeter (1955) and a Gould, Bahcall \& Flynn (1997)  for the
Maraston (1998) models.

\begin{deluxetable}{c c c c c c c c c}
\tablecaption{Observed and Derived Properties of M87 Globular Clusters\label{tab:res}}
\tablehead{
\colhead{Index} & \multicolumn{2}{c}{Metal-poor}& \multicolumn{2}{c}{Metal-rich} & \colhead{IMF\tablenotemark{a}} &
\multicolumn{3}{c}{Relative Age\tablenotemark{b} (Gyr)}\\
\cline{2-5}  \cline{7-9}\\
\colhead{} & \colhead{$m_{col}$}& \colhead{$m_{TO}$}& \colhead{$m_{col}$}& \colhead{$m_{TO}$}& \colhead{}
& \colhead{M98} & \colhead{W94} & \colhead{BC93} \\
}
\tablecolumns{9}
\tablewidth{0pt}
\tabletypesize{\scriptsize}
\startdata
$s_1$ & $0.67\pm0.015$ & $24.04\pm0.09$ & $1.27\pm0.017$ & $25.09\pm0.15$ & S55 & $1.3\pm3.1$ & $-1.9\pm3.0$ & $1.8\pm3.2$\\ 
 &  &  &  &  & MS79 & \nodata &$-1.3\pm2.3$ & $-0.5\pm2.2$ \\
 &  &  &  &  & GBF97 & $0.0\pm3.2$ & \nodata & \nodata \\
$s_2$ & $1.07\pm0.024$ & $23.99\pm0.09$ & $1.95\pm0.027$ & $25.00\pm0.14$ & S55 & $1.5\pm2.7$ & $-1.4\pm2.7$ & $2.5\pm2.9$\\ 
 &  &  &  &  & MS79 & \nodata & $-0.4\pm2.2$ & $-0.1\pm2.1$ \\
 &  &  &  &  & GBF97 & $0.4\pm2.7$ & \nodata & \nodata \\  
\enddata
\tablenotetext{a}{S55 = Salpeter 1955; MS79 = Miller \& Scalo 1979; GBF97 = Gould, Bahcall \& Flynn 1997.}
\tablenotetext{b}{M98 = Maraston 1998; W94 = Worthey 1994; MC93 = Bruzual \& Charlot 1993.}
\end{deluxetable}

The results are summarized in Table~\ref{tab:res}, where the reported
errors come from measurement uncertainties. We take the dispersion of the
age differences for each color index, $\sigma \sim 1.5$ Gyr, as an
indication of the systematic errors arising from the models.  As the
typical measurement error is 2-3 Gyr, and we have measurements using two
color indices, the mean of our results gives a formal age difference of
${\Delta}t \equiv {\rm age}_{blue} - {\rm age}_{red} = 0.2 \pm 1.5$
(systematic) $\pm 2$ (random) Gyr. Note that the quoted systematic
uncertainty does not include possible differences in turnover masses; a
10\% difference in the mass at the turnover would change the inferred age
difference by $\sim$ 1.5 Gyr.

\textit{Within the errors, all models are consistent with the two
populations being coeval}.  The average metallicities obtained from the
models for the subpopulations are [Fe/H]$_{blue}=-1.58$ and
[Fe/H]$_{red}=-0.30$, in reasonable agreement with the values
[Fe/H]$_{blue}=-1.41$ and [Fe/H]$_{red}=-0.23$ obtained by Kundu et~al.
(1999). Cohen et~al.  (1998) found no variation of age with metallicity
from their Keck spectra of M87 GCs, for which they find a mean age of
$13.2$ Gyr. Dividing the results presented in their Table 4 into two
populations and averaging the results gives a formal age difference of
$\Delta t \sim 1$ Gyr with a typical uncertainty in their age estimates of
$\sim 2$ Gyr. Our results are in good agreement with theirs, and both are
consistent with the subpopulations being coeval within the uncertainties.
This is gratifying as their sample of GCs is completely independent of
ours and they are using a different set of diagnostics from the models.
Indeed, as they determine absolute ages for each GC without making
additional assumptions about the GC masses, the agreement may be taken as
indirect evidence in support of our basic assumption that the GCLF
turnovers for the different subpopulations correspond to approximately the
same mass.

Although our observations were carried out in the same field as Kundu
et~al. (1999), we do not find support for their claim that the metal-rich
subpopulation is $3-6$ Gyr younger than the metal-poor.  As no uncertainty
is given for their estimate we are not able to assess the significance of
the disagreement, but none of the combinations of color indices, models
and IMFs that we have explored gives a result which falls within their
estimated range.

Our conclusions depend rather strongly on the assumption of equal mass at
the turnover for both subpopulations. Even though the agreement with Cohen
et~al.  (1998) and evidence from our Galaxy (McLaughlin \& Pudritz 1996)
suggest that this is a reasonable assumption, a direct test of this
hypothesis for some early-type galaxies is not only desirable but within
reach of efficient high-resolution spectrographs on $8$m-class telescopes.
It is also worth bearing in mind that the turnover for the metal-rich
population in the $u$ band falls near the level of $\sim 50\%$
completeness. This makes the use of parametric modeling (including the
completeness function) a necessity. Deeper observations will be required
to observe the metal-rich turnover directly.

As evidence accumulates for the GC subpopulations of undisturbed giant ellipticals being coeval
(Kissler-Patig et~al. 1998; Cohen et~al.  1998; Puzia et~al. 1999; Larsen et~al. 2002)
 models that rely on well separated epochs of star-formation, either via
major mergers or \textit{in-situ} bursts of star formation, appear less
favored as a \textit{generic} formation mechanism, unless one is willing
to push the separation of the two epochs to the point of being almost
coeval.  This is not to say that major mergers of late-type galaxies are
not a viable mechanism -- populations of young GCs have almost certainly
been identified in ongoing mergers (e.g. Whitmore \& Schweizer 1995) --
but their role as the primary mechanism for the formation of giant
ellipticals seems unlikely. Likewise, if two \textit{in-situ} star-forming
bursts are to be the formation mechanism for bright ellipticals, there has
to be enough time for the metals to diffuse and enrich the gas from which
the secondary GCs form. Lacking a specific mechanism which is responsible
for regulating this star formation, it is hard to constrain this idea.
Given the precision in the relative ages, the most we can say is that
enrichment, outflows/inflows and star formation process must have occured
within a $\sim$ few Gyr if this scenario is correct. Until a mechanism for
producing precisely two bursts is identified, this scenario amounts to the
\textit{assumption} that two metallicity populations imply two periods of
star formation.

Our results combined with those of Cohen et~al.  (1998) seem to show that
whatever the assembly history of these galaxies, the GCs associated with
M87 formed at at an early time, and within a short period. The presence of
two populations differing in metallicity by a factor of $\sim 20$, and yet
with roughly the same age, suggests that the origin of the two populations
may be related to differences in the local \textit{environments} where the
GC formed. According to the standard paradigm of structure formation,
galaxies form via accretion and merging of small objects. Simulations
predict a merger rate for massive cluster galaxies such as M87 that is
highly peaked at redshifts $z \gae 4$ (Gottl\"ober, Klypin \& Kravtsov
2001). This scenario provides the necessary difference in the local
environments as the GCs can form in protogalactic fragments of varying
mass. Simulations of the GC metallicity distributions based on this
picture have been succesful in reproducing the observations (C\^ot\'e,
West \& Marzke 2002).  As the bulk of star formation is expected to happen
early, and within $\sim$ a few Gyr, this picture for the formation of
bright ellipticals appears consistent with the existing observations of
their GC systems.

\acknowledgments

We are grateful to G. Bruzual, S. Charlot, C. Maraston and G. Worthey for
providing us with their models, and for help with their implementation.
Support for this work was provided by the National Science Foundation
through a grant from the Association of Universities for Research in
Astronomy, Inc., under NSF cooperative agreement AST-9613615 and by
Fundaci\'on Andes under project No.C-13442. Support for proposal \#09401
was provided by NASA through a grant from the Space Telescope Science
Institute, which is operated by the Association of Universities for
Research in Astronomy, Inc., under NASA contract NAS5-26555.

%
%

\end{document}